\documentstyle[eqsecnum,aps,prb,twocolumn,epsf,amssymb,floats]{revtex}

\newcommand{\PostScript}[3]{
  \vspace{#1cm}
\begin{center}
  \epsfysize=#2cm \leavevmode \epsfbox{#3}
\par
\end{center}
}

\begin{document}
\draft \preprint{TPR-99-06}

\title{Ionic and electronic structure of sodium clusters up
  to N=59}

\author{S. K\"ummel$^1$, M. Brack$^1$, and P.-G. Reinhard$^2$}
\address{$^1$Institute for Theoretical Physics, University of
  Regensburg, D-93040 Regensburg, Germany\\ e-mail:
  stephan.kuemmel/matthias.brack@physik.uni-regensburg.de\\ 
  $^2$Institute for Theoretical Physics, University of Erlangen,
  D-91077 Erlangen, Germany\\ e-mail:
  reinhard@theorie2.physik.uni-erlangen.de}

\date{To appear Sept.\ 15 in Phys.\ Rev.\ B {\bf 62}} \maketitle
\begin{abstract}
  We determined the ionic and electronic structure of sodium clusters
  with even electron numbers and 2 to 59 atoms in axially averaged and
  three-dimensional density functional calculations. A local,
  phenomenological pseudopotential that reproduces important bulk and
  atomic properties and facilitates structure calculations has been
  developed. Photoabsorption spectra have been calculated for
  $\mathrm{Na}_2$, $\mathrm{Na}_8$, and $\mathrm{Na}_9^+$ to
  $\mathrm{Na}_{59}^+$. The consistent inclusion of ionic structure
  considerably improves agreement with experiment. An icosahedral
  growth pattern is observed for $\mathrm{Na}_{19}^+$ to
  $\mathrm{Na}_{59}^+$. This finding is supported by photoabsorption
  data.

\end{abstract}

\pacs{PACS: 36.40.Vz,31.15.Ew,71.15.H,36.40.Wa}

\narrowtext \flushbottom

\section{Introduction}

Since the pioneering experiments of Knight {\it et al.}\cite{knight1}
and their interpretation in terms of the jellium
model,\cite{ekardt1,beck} sodium clusters have attracted great
attention, both experimentally and theoretically. This is due to the
fact that sodium is the ``simple metal'' par excellence, and thus is
best suited for the study of fundamental effects. From photoabsorption
experiments it is known that small sodium clusters have overall shapes
that strongly vary with size and are determined by electronic shell
effects.\cite{ep,selby,borggreen1,sajm,manninen1} On the other
hand, cold clusters with several thousand atoms build
icosahedra,\cite{tpmartin} i.e., they show ionic shell effects,
whereas the bulk material crystallizes in a bcc lattice. Recent
experiments indicate that both ionic and electronic degrees of freedom
play a role in determining the structural and thermal properties of
clusters with several tens of atoms.\cite{schmelzen}

On the theoretical side, there have been different approaches to
obtain an understanding of the delicate interplay between ions and
electrons that is the source of this variety. On one hand,
relatively transparent models like the shell and jellium models in
several levels of sophistication have been widely
used.\cite{ekardt1,beck,shellmodel,ultimatejellium,bertschjel,revmod,guetxc,alasia}
On the other hand, quantum chemistry and density functional theory
offer methods to study clusters on the highest level of sophistication
presently possible.  But {\it ab initio} calculations in the strictest
sense, i.e., taking all electrons into account, have
only been performed for a few selected cases for the smallest
clusters, due to the enormous computational effort that they
require.\cite{kouteckyneu1,kouteckyrev,guan} Taking only the valence
electrons into account reduces the complexity considerably, but the
expense for searching low-energy configurations in three dimensions
\cite{martins,moullet,roethlis,koutecky96,rubio} still grows rapidly
with the system size. The largest {\it ab initio} studies of sodium
clusters to our knowledge are the recent finite-temperature
investigations presented in Refs. \onlinecite{rytkoenen,akola00}.

Several models have been developed to bridge the gap between the {\it
  ab initio} calculations and the shell and jellium models.  The
"Spherically Averaged Pseudopotential Scheme" (SAPS) describes the
ions by pseudopotentials, in most cases local ones, and the valence
electrons are restricted to spherical symmetry.\cite{saps} Models
based on a volume-averaged or perturbational treatment of ionic
effects\cite{alonso,ppstoer} improve on the treatment of the
electrons. Yet 
further approaches are the H\"uckel and related
models,\cite{spiegelmann1,spiegelmann2} molecular
dynamics based on empirical potentials,\cite{calvo} and
recently, also the extended Thomas-Fermi model combined with a local
pseudopotential has been used to study sodium clusters.\cite{guettf,aguado}

From the above examples, two points become clear that considerably facilitate
systematical studies of clusters with more than twenty atoms. First, the fact
that many calculations make use of phenomenological local pseudopotentials
shows the importance of such potentials. This is especially true for sodium,
where previous investigations\cite{alasia,evc,guetli} have shown that a local
electron-ion interaction can be a good approximation. But especially in cluster
physics, where one bridges the region between the atom and the bulk, it is
important that a pseudopotential give reliable results for all sizes despite
its locality.  Second, one needs models that make it computationally
manageable to calculate structures and optical properties of clusters
systematically for a wide range of considerable sizes, but which on the other
hand leave the underlying physics intact and are detailed enough so that
relevant information can be drawn from them. Besides extending the
computational range, such models will serve the even more important purpose to
distill the dominant physical effects from the wealth of details that fully
{\it ab initio} calculations supply. The results of the model calculations of
course must be verified in calculations of higher accuracy and by comparison
with experiment. These tasks are addressed in the present article. 

In Section \ref{pp} we develop a new phenomenological pseudopotential which
meets the just mentioned requirements. Detailed comparisons with {\it ab
  initio} calculations for the smallest clusters in section \ref{tests} verify
the validity of the pseudopotential and an axially averaged density
functional model\cite{caps} for structure calculations. In section
\ref{structures}, a systematic survey over  
structures and photoabsorption spectra of Na clusters up to
$\mathrm{Na}_{59}^+$ is given. The results of the survey are
summarized and discussed in the concluding section \ref{conclude}.

\section{A local pseudopotential for structure calculations}
\label{pp}

Rigorous pseudopotentials in the sense of Phillips
and Kleinman \cite{phillips} and modern {\it ab initio}
pseudopotentials\cite{bachelet,troulliermartins} are always nonlocal.
However, it has been noted early in the development of pseudopotential
theory that by relaxing the Phillips-Kleinman condition, one can open up a new
class of pseudopotentials.\cite{heineaba} These are 
also termed ``model potentials'' because they are constructed by
choosing some analytical functions as models for the partial-wave
potentials and adjusting their parameters such that some
experimentally known quantities, e.g., atomic energy levels, are
matched.  If several partial-wave potentials can be chosen to be the
same, one can construct a local model potential. In
fact, such local phenomenological potentials have been and are being used
successfully in many branches of physics, see, e.g., Refs.\ 
\onlinecite{ppstoer,aguado,evc,popovic,landman}. However, the local
approach can 
only be expected to be good for the so-called ``nearly free electron''
metals whose atomic structure consists of only $s$ or $p$ electrons
outside of an ionic core with a noble gas configuration.  (Lithium,
with its lack of $p$ electrons in the core, is the prominent
exception.)  In these metals, the two contributions making up the
pseudopotential have a tendency to cancel each other, and the effective
potential 
a valence electron experiences is further diminished by screening
effects.  Therefore, model potentials have been proposed in the past
which tried to exploit this simple electronic structure either by
fitting properties of the bulk solid,\cite{evc,caps,popovic,emptycore}
or of the single atom,\cite{heineaba} or by {\it ad hoc}
fulfilling desired numerical properties.\cite{guettf,calvayrac}
However, some of these potentials can lead to wrong predictions when
they are used in physical surroundings different from the one in which
they were set up,\cite{prbrc} or when properties other than the
adjusted ones are looked at. With an emphasis on solid state
properties, the last point has been discussed in detail previously,
see, e.g., Ref.\ \onlinecite{evc} for an overview. Our aim here
is to develop a local model potential that gives a maximum degree of
transferability in the sense that potentials constructed according to
our scheme should reproduce the important physical properties of a
system, irrespectively of its number of atoms or the way in which these
are arranged.

\subsection{The Ansatz}

The construction of a model potential consists of two steps. The first
is to choose the model function. It should allow for a
physical interpretation of the final potential,  
and at the same time should have analytical and numerical properties
that allow for an easy application. The parameterization 
\begin{equation}
  V_{\rm ps}\left(r\right) = -\frac{Ze^2}{r}\left\{
    c_1\mbox{erf}\left(\frac{r}{\sqrt{2}\sigma_1}\right) +
    c_2\mbox{erf}\left(\frac{r}{\sqrt{2}\sigma_2}\right) \right\}
  \quad,
\label{defpp}
\end{equation}
where
\begin{equation}
  \mbox{erf}(x) = \frac{2}{\sqrt{\pi}}\int_0^x dy\,\exp{(-y^2)},
\label{eq:erf}
\end{equation}
certainly meets the second requirement, since the error function can
very efficiently be handled numerically\cite{numrec} and yields a
smooth representation on a grid. That this parameterization also has a
transparent physical interpretation will be demonstrated at the end of
this section. The second step is the choice of the four parameters
$\sigma_1,\sigma_2,c_1,c_2$. One necessary condition is that the
correct Coulomb limit,
\begin{equation}
  \lim_{r\rightarrow\infty} V_{\rm ps}\left(r\right)=\frac{-Ze^2}{r},
\end{equation}
be obtained, which requires
\begin{equation}
  c_1+c_2=1.
\label{c1c2}
\end{equation}
Thus we are left with three free parameters. Since our aim is to
construct a pseudopotential that gives reliable properties for all
clusters, i.e., spanning the region from the atom to the bulk, the
most important properties of both atom and bulk solid must be
reproduced.  Therefore, we have chosen to fit the Wigner-Seitz radius
$r_{\mathrm s}$ and the compressibility $B$ of the crystalline metal,
together with the energy of the atomic $3s$ level $e_{\mathrm a}$. These
quantities characterize sodium and influence both structure and
electronic excitations, thus being of great importance for reliable
results.  A test for whether we really have captured the relevant
physics will be to check if non-fitted quantities (bond-lengths,
atomic spectra, bulk binding, dipole resonances) are also reproduced
correctly.

\subsection{Determination of basic properties}

We calculate the bulk properties in second-order perturbation
theory.\cite{evc} The unperturbed system is the
noninteracting homogeneous gas of valence electrons, the perturbation
is given by the potentials arising from the crystal lattice and from
the interaction of the electrons with each other. Each crystal ion is
described by a pseudopotential $V_{\mathrm ps}$ centered on a lattice
site. Up to second order, the binding energy per valence electron
$e_{\mathrm b}$ is given by
\begin{eqnarray}
\label{ebind}
e_{\mathrm b}(r_{\mathrm s})&=& e_{\mathrm kin}(r_{\mathrm s}) +
e_{\mathrm xc}(r_{\mathrm s}) + e_{\mathrm ps1}(r_{\mathrm s}) +
e_{\mathrm h}(r_{\mathrm s}) + e_{\mathrm ii} \nonumber \\ && +
e_{\mathrm bs}(r_{\mathrm s}).
\end{eqnarray}
Here $ r_{\mathrm s}=\left[3/\!\left(4\pi n\right)\right]^\frac{1}{3}
$ is defined in terms of the average valence electron density $n$,
\begin{equation}
  e_{\mathrm kin}(r_{\mathrm s})=\frac{3 \hbar^2}{10 m}\left(\frac{9
      \pi}{4}\right)^\frac{2}{3} \frac{1}{r_{\mathrm s}^2}
\end{equation}
is the noninteracting kinetic energy, and $e_{\mathrm xc}(r_{\mathrm
  s})$ is the exchange and correlation energy for which we have
employed the LDA functional of Perdew and Wang.\cite{pw} The quantity
\begin{equation}
  e_{\mathrm ps1}(r_{\mathrm s})=\frac{3}{Z 4\pi {r_{\mathrm
        s}}^3}\int V_{\mathrm ps}(r)d^3r
\label{1storder}
\end{equation}
is the first-order contribution of the pseudopotential to the binding
energy, $e_{\mathrm h}(r_{\mathrm s})$ is the Hartree energy of the
valence electron density, and $e_{\mathrm ii}$ is the electrostatic
energy of point ions. Finally, $e_{\mathrm bs}(r_{\mathrm s})$ is the band
structure energy discussed below. Integrations are taken over the (infinite)
volume of the crystal. These contributions are rewritten as
\begin{eqnarray}
  \lefteqn{e_{\mathrm ps1}(r_{\mathrm s}) + e_{\mathrm h}(r_{\mathrm
      s})+ e_{\mathrm ii}(r_{\mathrm s}) =} \nonumber \\ && \frac{3}{Z
    4 \pi r_{\mathrm s}^3} \int \left( V_{\mathrm ps}+\frac{Z e^2}{r}
  \right)d^3 r+\nonumber \\ && \frac{3}{Z 4 \pi r_{\mathrm s}^3} \int
  \left( \frac{V_{\mathrm h}}{2}-\frac{Z e^2}{r} \right)d^3 r
  +e_{\mathrm ii}
\end{eqnarray}
to obtain the volume-averaged repulsive part of the
pseudopotential that defines its strength
\begin{equation}
\label{strength}
{\cal S}_{\mathrm ps}= \int \left( V_{\mathrm ps}+\frac{Z e^2}{r}
\right)d^3 r,
\end{equation}
and the total electrostatic energy of point ions in a compensating
uniform negative background, called the Madelung energy $e_{\mathrm
  m}$.  Separating the Coulomb force into long and short-range
components,\cite{hafner} one can calculate $e_{\mathrm m}$, and for
the bcc lattice one obtains $e_{\mathrm m}=-0.895929 Z^\frac{2}{3}
e^2/r_{\mathrm s}$.  Thus the only characteristic\cite{stabiljel} of
the pseudopotential that enters the binding energy in first-order
perturbation theory is its strength $\cal{S}_{\mathrm ps}$. For fixing
the radial dependence of the potential it is therefore essential to
take into account the second-order band structure energy
\begin{equation}
  e_{\mathrm bs}(r_{\mathrm s})=\frac{1}{2} \frac{3}{Z^2 4 \pi r_{\mathrm s}^3}
  \sum_{{\bf q \ne 0}} \left| \tilde{V}_{\mathrm ps}\left({\bf
        q}\right) S\left({\bf q}\right) \right|^2 \frac{\chi\left(
      {\bf q}\right)}{\epsilon\left( {\bf q}\right)}.
\label{ebs}
\end{equation}
In (\ref{ebs}),
\begin{equation}
  S\left({\bf q}\right)=\frac{1}{N} \sum_{j=1}^N \exp\left( -i {\bf q
      R}_j \right)
\end{equation}
is the structure factor with the sum running over all ionic positions.
For a lattice without basis one obtains $S\left({\bf
    q}\right)=\delta_{\bf q,K}$, where $\bf K$ denotes reciprocal
lattice vectors, i.e., the sum in (\ref{ebs}) is running over
reciprocal lattice vectors only. Furthermore,
\begin{equation}
  \chi\left( {\bf q}\right)=-\frac{m k_F}{2 \pi^2 \hbar^2}\left(
    1+\frac{1-\eta^2}{2\eta}\ln\left| \frac{1+\eta}{1-\eta}
    \right|\right), \hspace{2mm} \eta=\frac{q}{2k_F}
\end{equation}
is the Lindhard function with $k_F=(3 \pi^2 n)^\frac{1}{3}$, and
\begin{equation}
\label{dielectricf}
\epsilon\left( {\bf q}\right)=1-\frac{4 \pi e^2}{q^2}\left(
  1-\mathcal{G}\left( {\bf q}\right) \right) \chi\left( {\bf q}\right)
\end{equation}
is the dielectric function including the local field
correction
\begin{equation}
  {\mathcal G}=-\frac{q^2}{4\pi e^2}\frac{d^2}{dn^2}\left(n e_{\mathrm
      xc}\left[n\right]\right).
\end{equation}
in the LDA.  The pseudopotential enters via its Fourier transform,
which for our potential is given by
\begin{equation}
  \tilde{V}_{\mathrm ps}\left({\bf q}\right)= -\frac{4\pi e^2 Z}{q^2}
  \sum_{i=1}^2 c_i\exp\left(-\frac{q^2 \sigma_i^2}{2}\right).
\end{equation}
From Eqs.\ (\ref{ebind}) to (\ref{dielectricf}) we calculate
$e_{\mathrm b}$; the minimum of $e_{\mathrm b}$ determines $r_{\mathrm
  s}$, and the bulk modulus is obtained from
\begin{equation}
  B=-V \frac{\partial P}{\partial V}=\frac{1}{12 \pi r_{\mathrm
      s}}\left( \frac{\partial^2 e_{\mathrm b}}{\partial^2 r_{\mathrm
        s}}- \frac{2}{r_{\mathrm s}}\frac{\partial e_{\mathrm
        b}}{\partial r_{\mathrm s}} \right).
\end{equation}
The reciprocal lattice vectors are generated numerically from
the reciprocal basis, and we carefully checked that in all
calculations the sum over reciprocal lattice vectors was numerically
converged. 

The atomic $3s$ level is the lowest eigenvalue $e_{\mathrm a}$ of the
Schr\"odinger equation
\begin{equation}
\label{atom}
\left( -\frac{\partial^2}{\partial r^2} +\frac{l\left(l+1\right)}{r^2}
  +\frac{2 m}{\hbar^2}V_{\mathrm ps}\left(r\right) -\frac{2
    m}{\hbar^2}e_{\mathrm a} \right)u\left(r\right)=0
\end{equation} 
for $l=0$, where the pseudo wavefunction has, as usual, been factorized into
radial and angular components, $
\psi(r)=\left[u\left(r\right)\!/r\right] \, Y_{lm}(\vartheta,\varphi)
$. At this point,  a subtlety should be
considered. The atomic energy calculated within density functional
theory using the LDA will slightly differ from  $e_{\mathrm a}$ due to
the well known fact that in the LDA, the self interactions in the Hartree and 
exchange energy do not cancel each other exactly and leave a spurious self
interaction 
energy.\cite{psic1} One therefore might be tempted to ``compensate''
this self-interaction energy by adjusting the pseudopotential
parameters such that the 
experimental value is matched when the self-interaction energy is
included. This, however, would be dangerous for several reasons. First, it
must be recalled that for an accurate description of bonding properties, the
pseudopotential must lead to an accurate description of 
the electronic density. If one adjusts the pseudopotential parameters such
that a spurious energy is compensated, then the valence electron density
resulting from this pseudopotential might correspondingly show spurious,
unphysical deformations, leading to wrong bonding properties, as
discussed previously.\cite{cohen} Secondly, the self-interaction
energy becomes 
smaller with increasing delocalization of the electrons. For an
electron delocalized over a volume $\Omega$, the self-interaction correction
is of order\cite{psic2} $\Omega^{-\frac{1}{3}}$. Since the
valence electrons in metal clusters are delocalized to a high degree, and
since the cluster volume changes noticeably in the size range from N=8 to N=58
that we are interested in, a compensation of the atomic self
interaction energy via the  
pseudopotential would lead to problems for increasing cluster sizes. And
third, building the self-interaction energy into the 
pseudopotential would be inconsistent with the procedure of
fitting to the bulk, because the bulk calculation is less affected by the
self-interaction error due to the delocalization of the bulk
electrons. Therefore, it is better to exactly solve Eq.\ (\ref{atom}). 
This can straightforwardly be done numerically by combining the
Runge-Kutta method with adaptive stepsize control with a globally
convergent Newton scheme. As a welcomed side effect, our
pseudopotential might also be usable in self-interaction free, i.e.,
beyond-LDA density functional calculations.

\subsection{Results and comparison}

The three free parameters of our model potential (\ref{defpp}) were
now chosen such that the experimental low-temperature value\cite{am}
$r_{\mathrm s}=3.93 a_0$ be reproduced exactly, while at the same time
$e_{\mathrm a}$ and $B$ be as close as possible to the experimental
values\cite{heineaba} $e_{\mathrm a}=-0.378 Ry$ and\cite{compress} $B=0.073
Mbar$. These conditions lead to the parameters
\begin{eqnarray}
\label{parameters}
\sigma_1=0.681 \, \rm{ a_0}, && \hspace{0.5cm} c_1=-2.292,\nonumber \\ 
\sigma_2=1.163 \, \rm{ a_0}, && \hspace{0.5cm} c_2=3.292.
\end{eqnarray}
In Table \ref{comppp} we have listed the resulting $r_{\mathrm s}$,
$B$ and $e_{\mathrm a}$ for our smooth-core pseudopotential and for other local
pseudopotentials that have widely been used in cluster physics.
\begin{table}[h]
\begin{tabular}{|c|d|d|d} 
  Pseudopotential& $r_{\mathrm s}/a_0$ & $B/Mbar$ & $ e_{\mathrm a}/eV$ \\ 
  \hline 
  Empty core, rc= 1.66$^a$ & 3.49 & 0.119& -5.52 \\ \hline
  Empty core, rc= 1.76$^b$ & 3.61& 0.109& -5.32 \\ \hline 
  Ref.\ \onlinecite{caps} & 3.84 & 0.079 & -5.31 \\ \hline
  Heine-Abarenkov$^c$ & 3.90& 0.080& -5.12 \\ \hline 
  Present work & 3.93 & 0.074 & -5.18 \\ \hline 
  Experiment & 3.93 & 0.073 & -5.14 \\ 
\end{tabular}
\caption{Equilibrium density $r_{\mathrm s}$, bulk modulus $B$, and atomic $3s$
  level $e_{\mathrm a}$, calculated for standard local pseudopotentials: (a)
  Ref.\ \protect \onlinecite{emptycore}, (b) Ref.\ \protect
  \onlinecite{saps}, and (c) Ref.\ \protect
  \onlinecite{popovic,landman,ppstoer}. Bottom line: Experimental results. 
  See text for details and references for experimental values.}
\label{comppp}
\end{table}  
The empty-core potential with both of the most frequently used choices
for its cut-off radius $r_c$, and the pseudopotential of Ref.\ 
\onlinecite{caps} that was constructed in the same spirit and adjusted
in first-order perturbation theory only, lead to considerable
deviations from the experimental values for all quantities.  The local
Heine-Abarenkov potential\cite{ppstoer,popovic,landman} gives a
reasonable $r_{\mathrm s}$ and an $e_{\mathrm a}$ very close to the
experimental one, but 10\% error in $B$. Our pseudopotential by
construction gives no or only very small differences from the
experimental values for $r_{\mathrm s}$, $B$ and $e_{\mathrm a}$, showing that
it is possible to obtain reasonable results for all these quantities
with a local potential. However, a severe test will be whether also
non-fitted quantities are reproduced correctly. To this end, we have
calculated the bulk binding energy $e_{\mathrm b}$, the dimer binding
length $d_{\mathrm dimer}$, the zero of the pseudopotential form
factor, given by
$q_0=\sqrt{2[\ln(c_2)-\ln(-c_1)]/(\sigma_2^2-\sigma_1^2)}$ in our
case, and the first seven excited atomic energy levels.  Table
\ref{nonfitted} shows that also these ten non-fitted quantities come
out very close to the measured values, revealing that the model
potential defined by (\ref{defpp}), (\ref{parameters}) indeed
incorporates the relevant physical effects.

\begin{table}[h]
\begin{tabular}{|c|d|d|}
  & Present work & Experiment \\ \hline 
$ e_{\mathrm b}/eV$ & -6.19 & -6.25 \\ \hline 
$q_0/(2k_F)$ & 0.92 & 0.87$^a$/0.98$^b$ \\ \hline
$d_{\mathrm dimer}/a_0$ & 5.78 & 5.82 \\ \hline 
3s - 3p $/eV$ & 2.19  & 2.10 \\ \hline 
3s - 4s $/eV$ & 3.18 & 3.19 \\ \hline 
3s - 3d $/eV$ & 3.68 & 3.52 \\ \hline 
3s - 4p $/eV$ & 3.80 & 3.75 \\ \hline 
3s - 5s $/eV$ & 4.13 & 4.12 \\ \hline 
3s - 4d $/eV$ & 4.33 & 4.29 \\ \hline 
3s - 5p $/eV$ & 4.39 & 4.35
\end{tabular}
\caption{Left column:  Bulk binding energy $e_{\mathrm b}$ from perturbation 
  theory; normalized zero of the form factor $q_0/(2k_F)$; dimer
  binding length $d_{\mathrm dimer}$ calculated with our smooth-core
  pseudopotential and CAPS; valence electron excitation energies for
  our pseudopotential.  Right column: Measured bulk binding
  energy;\cite{emsley} semi-empirical values for $q_0/(2k_F)$ from
  optical properties (a) and from elastic constants (b), see Ref.\ 
  \protect \onlinecite{expQ0} for a discussion; dimer binding
  length;\cite{expdimer} and energy level differences as obtained from
  spectroscopic lines.\cite{hakenwolf}}
\label{nonfitted} 
\end{table}

\begin{figure}[htb]
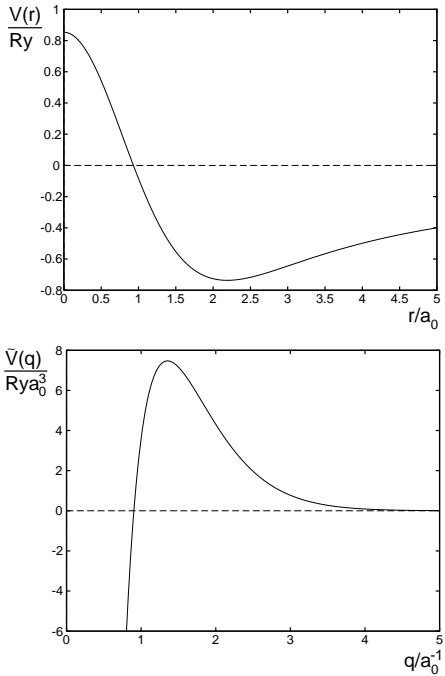

  \PostScript{0}{9}{ppplots.epsi}
\caption{The smooth-core pseudopotential in real and Fourier space.}
\label{ppplots}
\end{figure}
Fig.\ \ref{ppplots} shows our pseudopotential in real and in Fourier
space, leading us back to the question of the motivation for the
ansatz (\ref{defpp}). From Eq.\ (\ref{c1c2}), one can interpret the
coefficients $c_1$ and $c_2$ as the charges associated with the
attractive and repulsive terms, respectively, in the pseudopotential.
This interpretation becomes more transparent when one looks at the
corresponding pseudocharge density. The pseudocharge $n_{\rm ps}$ is
related to the pseudopotential via Poisson's equation
\begin{equation}
\label{poisson}
\triangle V_{\rm ps}\left(r\right) = 4 \pi e^2 n_{\rm
  ps}\left(r\right),
\end{equation}
and for our pseudopotential is given by
\begin{equation}
  n_{\mathrm ps}(r)= n_1\exp\left(-\frac{r^2}{2{\sigma_1}^2}\right)+
  n_2\exp\left(-\frac{r^2}{2{\sigma_2}^2}\right),
\label{pden}
\end{equation}
where
\begin{equation}
  n_i=c_i {\left(2\pi{\sigma_i}^2\right)}^{-\frac{3}{2}} \quad
  {\mathrm for} \quad i=1,2.
\end{equation}
Thus, our parameterization describes the ionic core by two overlapping
Gaussian charge densities. The different heights and widths of the
Gaussians result in a pseudodensity that is negative for small
distances, corresponding to a repulsive core, and has a positive tail
that compensates the core for larger distances. Equation (\ref{pden})
thus is a generalization of the parameterization that was used in our
previous studies,\cite{prbrc} where the pseudodensity has a two
step-profile. Besides its transparent physical interpretation, the
pseudopotential (\ref{defpp}) has excellent numerical properties: in
real space the potential is smooth with no extreme peaks and can very
efficiently be handled numerically via the pseudodensity, allowing to
solve the Coulomb problem for electrons and ions together.  At the
same time, its rapidly converging Fourier transform makes the
potential equally well suited for calculations in reciprocal space.

\section{Comparison with ab initio results}
\label{tests}

In the present work we have employed the
``Cylindrically Averaged Pseudopotential Scheme'' (CAPS), which has
been introduced earlier\cite{caps,prbrc} and which has been modified
and improved for the present purposes. Since CAPS treats the
electronic degrees of freedom in the Kohn-Sham formalism without
limiting them to spherical symmetry, explicitly includes the ionic
structure, and furthermore is numerically efficient, it is a very good
tool for systematical studies of the interplay between ions and
electrons. But before CAPS and the smooth-core pseudopotential are
applied on a large scale, we further test their usefulness for cluster
structure calculations.

\begin{figure}[h]
\PostScript{0}{8}{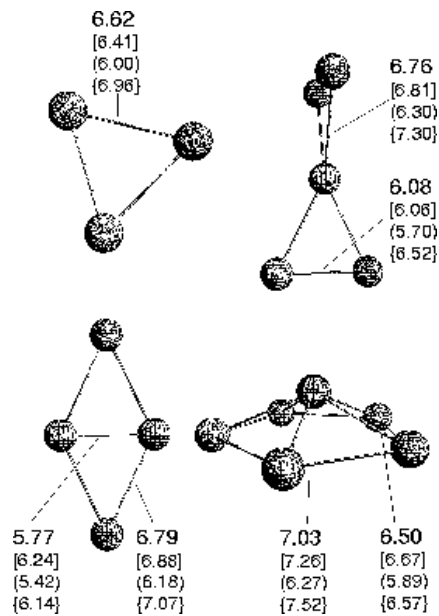}
\caption{Ground state structures for $\mathrm{Na}_{3}^+$, $\mathrm{Na}_{5}^+$,
  $\mathrm{Na}_{4}$ and $\mathrm{Na}_{6}$ as obtained in CAPS with the smooth
  core pseudopotential. The numbers report the bond lengths in $a_0$, where
  the uppermost values in each column are from the present work, the ones
  in square brackets from Hartree-Fock/CI
  calculations,\cite{koutecky96,kouteckyneu1} the ones in parentheses
  from DFT with LDA,\cite{martins,moullet} and the ones in braces from
  all-electron Hartree-Fock calculations.\cite{kouteckyneu1,kouteckyc}} 
\label{compabini}
\end{figure}

That the smooth core pseudopotential reproduces the experimentally
known dimer bond length has already been shown in Table
\ref{nonfitted}. For the next larger clusters, no direct experimental
information on structures or bond lengths is available, and we
therefore resort to comparisons with other theoretical work. Fig.\
\ref{compabini} shows the four smallest Na clusters with even electron
numbers as they are obtained in CAPS with the smooth core
pseudopotential. For these clusters, also three-dimensional geometry
optimizations have been performed using {\it ab initio}
pseudopotentials and Hartree-Fock with configuration interaction (CI)
\cite{kouteckyneu1,koutecky96}, or DFT with the LDA
\cite{martins,moullet} for the valence electrons, respectively. Also
all-electron Hartree-Fock calculations have been
reported.\cite{kouteckyneu1,kouteckyc} CAPS finds exactly the same
structures as the three-dimensional methods, and this in spite of the
fact that the ionic configurations of $\mathrm{Na}_{4}$
and
$\mathrm{Na}_{5}^+$ might let the cylindrical averaging seem a rather
far-fetched approximation. But not only the overall geometry, also
the bond lengths are in good agreement with the ones from the
three-dimensional calculations. Whereas all-electron Hartree-Fock
overestimates bond lengths due to missing correlation effects, the
three-dimensional LDA calculations lead to the well-known
underestimation. By construction, the phenomenological pseudopotential
compensates this underestimation, and it is thus reassuring to see
that the resulting bond lengths are close to the ones found in the
quantum chemistry calculations. Three-dimensional calculations
have been performed for a few other neutral clusters,\cite{roethlis}
and CAPS is in agreement with the three-dimensional geometries in
these cases, too. A detailed analysis of structures of neutral
clusters and their static electric polarizability can be found in Ref.\
\onlinecite{epjd}. 

\begin{figure}[bht]
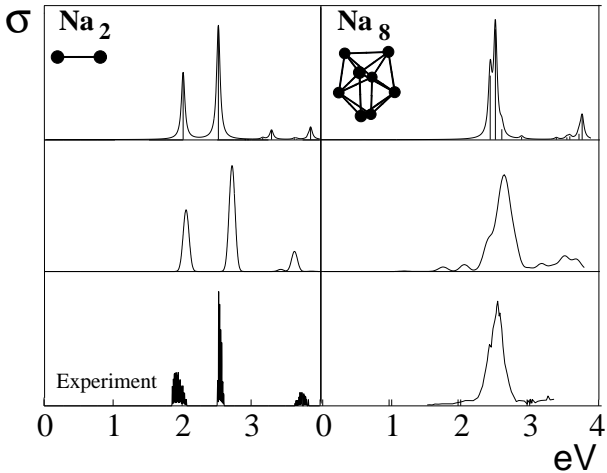

\PostScript{0}{6.5}{na4u8comp.epsi}
\caption{Photoabsorption cross section $\sigma$ in arbitrary units against
  energy in eV for $\mathrm{Na}_2$ and $\mathrm{Na}_8$. From top to
  bottom: present results for the CAPS structure shown in the
  inset, TDLDA calculations for the corresponding three-dimensional structures,
  \cite{chelikowsky} experimental spectra.\cite{neupabs2,neupabs1} The TDLDA
  and collective model results have been broadened by 0.06 eV to simulate the
  finite line width of the experiment.}
\label{na4u8comp}
\end{figure}

A further test for structure calculations is obtained by comparing the
photoabsorption spectra corresponding to the theoretically determined
structures to the measured ones. To this end, we have calculated the
valence electron excitations for our cluster structures in a
collective approach. The basic idea of this method has been presented
earlier,\cite{brack1,lrpa} and a detailed discussion of its extensions and the
relation to density functional theory is the subject of a forthcoming
publication.\cite{tobepucoll} Here, we demonstrate the validity of our
collective model for the test cases $\mathrm{Na}_2$ and
$\mathrm{Na}_8$, where experimental data and recent {\it ab initio}
TDLDA calculations are available for comparison. The first row of
plots in Fig.\ \ref{na4u8comp} shows the excitation  
spectra obtained with the axial collective model for the CAPS
structures
shown in the insets, the second row shows the TDLDA spectra
for the corresponding three-dimensional geometries, and the bottom row
shows the experiment.\cite{neupabs2,neupabs1} The lower two rows were
directly adapted from Ref.\ \onlinecite{chelikowsky}, and the collective model
results were then plotted on the same scale. Besides small differences
in the two small peaks seen at high energies for $\mathrm{Na}_2$, the
agreement between the three sets of data is very good. This
demonstrates that our collective model is capable of correctly describing
excitations even in small systems. It further is to be noticed
that the excitation energies obtained with the smooth-core
pseudopotential are $\approx$ 0.15 eV lower than the TDLDA energies and
closer to the experimental ones, which can be attributed to the larger
bond lengths that result with the present pseudopotential.

\section{Structures and photoabsorption spectra of singly-charged
  cationic sodium clusters} 
\label{structures}

Since the comparisons in the previous section have shown that the
phenomenological pseudopotential, the collective model and CAPS are well
suited for the description of sodium clusters, a systematic study of clusters
with even electron numbers between 8 and 58 is presented. For each cluster a
great number of simulated annealing runs was started from different random
configurations. The search was continued until new runs no longer returned new
low-energy structures. Although it must always be kept in mind that no
practical algorithm guarantees that all low-energy structures are found, this
extensive and unbiased procedure at least gives good hope to do so. The aim of
this survey is to investigate how electronic and ionic shell effects play
together in determining the cluster structure, and thus gain a better
understanding of how matter grows. The geometry optimization was done with
CAPS, but the energies of the resulting structures have also been
obtained in three-dimensional Kohn-Sham calculations to check the
ordering of isomers without axial 
restriction. Photoabsorption spectra are calculated with the collective
model, and comparison with the experimental spectra gives further information
on the relevant structures.

\begin{figure}[h]
\begin{center}
\PostScript{0}{5}{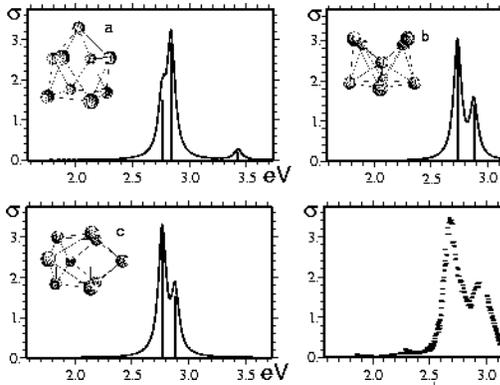}
\caption{Low-energy configurations for 
  $\mathrm{Na}_{9}^+$, and corresponding photoabsorption spectra from
  the collective model. A phenomenological Lorentzian line broadening
  of width 0.08 eV has been applied. Dotted curve: experimental cross
  section for T=105 K.\cite{expcoldclosed,haberlandepjd}}
\label{na9pall}
\end{center}
\end{figure}

The ``magic'' cluster $\mathrm{Na}_{9}^+$ is spherical according to the
jellium model, and the ground state found with CAPS is the $C_{4v}$ structure
(a) in Fig.\ \ref{na9pall}. Separated by 0.10 eV, CAPS finds the $D_{4d}$
isomer (b), and both these geometries were also found in three-dimensional CI
and LDA calculations.\cite{kouteckyrev,calvayrac}  In addition, CAPS finds
the third isomer (c) which is higher by 0.23 eV. All
these clusters have nearly spherical valence electron densities, and therefore
support the shell model picture. But the ionic configurations are, of course,
non-spherical, and this is reflected in the photoabsorption cross sections.
The first thing to be noted is the overall position of the resonance, about
which there has been a long-standing
debate.\cite{ekardt1,revmod,guetxc,rubio,ppstoer,bertsch,isspic} In
previous work\cite{prbrc} we have shown that when ionic structure is
taken into account
via a pseudopotential, the detailed form of the potential greatly influences
the resonance position. But whereas our previous calculations were mainly
compared to high-temperature data, and, due to limited collective model basis
sets, fixed the resonance positions only within a few percent (as pointed out
in Ref.\ \onlinecite{prbrc}), we have now fully converged the basis. The
resulting overall resonance positions are very close to
the experimental low-temperature positions, showing that consistent
inclusion of ionic structure gets the plasmon
position in the correct energy interval. The optical response of
$\mathrm{Na}_{9}^+$ has also been theoretically investigated
previously,\cite{kouteckyrev,koutecky96,ppstoer,calvayrac} and each of these
calculations explains the experimental spectrum on the basis of a different
isomer. In our calculations, the spectra of isomers (b) and (c) both reproduce
the splitting of the main peak seen in the experiment. Isomer (a) does not
show this feature, but it leads to the small sub-peak around 3.5 eV that is
also seen experimentally. The situation becomes transparent when the results
from our three-dimensional calculations, see Table \ref{allen}, are taken into
account. They show that isomer (c) is degenerate with isomer (a), and the
experimental spectrum can thus be explained even at low temperatures as
resulting from a mixture of isomers (a) and (c). (The importance of
isomerism on dipole spectra in small sodium clusters has also been
pointed out in Ref.\ \onlinecite{alasia2}.)

\begin{figure}[thb]
  \PostScript{0}{3.5}{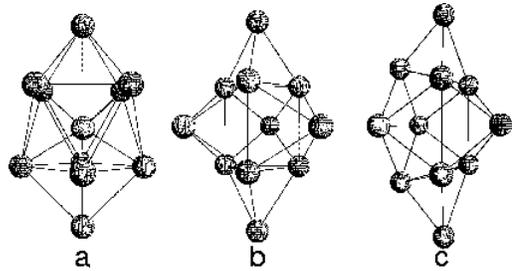}
\caption{Low-energy configurations for $\mathrm Na_{11}^+$. See text for a
  discussion of photoabsorption spectra.}
\label{struct11pall}
\end{figure}

The lowest energy structure that we find for $\mathrm{Na}_{11}^+$ ($D_{4d}$,
labeled (a) in Fig.\ \ref{struct11pall}) can be understood as resulting from
placing one atom in the center of $\mathrm{Na}_{10}$, or capping isomer (b) of
$\mathrm{Na}_{9}^+$ on the quadratic faces.  Separated by 0.11 eV and 0.19 eV,
respectively, CAPS finds the $D_{3h}$ isomer (b), and its deformed counterpart
(c), which is only a shallow local minimum in CAPS and easily
transforms into the more stable structure (b). In the three-dimensional
calculations, the ordering of isomers is reversed: (b) becomes the
ground state,
 and our results are thus consistent with the findings of Ref.\
\onlinecite{koutecky96}. We have calculated the photoabsorption spectra with
the collective model and obtain two resonances with heights approximately
1:2 for all three geometries. These are centered at 2.41/2.94 eV for structure
(a), at 2.39/2.94 eV for structure (b), and at 2.27/2.98 eV for structure
(c). The experimental photoabsorption spectrum \cite{expcoldopen} of 
$\mathrm{Na}_{11}^+$ at 380 K shows two broad peaks with heights 1:2 around 
2.2 eV and
2.85 eV, and a pattern of six peaks when measured at 35 K: three low intensity
ones at 1.9 eV, 2.2 eV and 2.4 eV, and three high intensity ones at 2.5 eV,
2.8 eV and 3.0 eV. Obviously, the calculated resonances are blue shifted by a
few percent with respect to the hot experiment. This is understandable since
the calculations have been done for T=0 K, and thermal expansion of the
cluster shifts the plasmon to slightly lower energies. (This effect has
recently been put into evidence quantitatively for the static
response.\cite{expandpol}) When compared to the cold data, the calculated
resonances are in the correct energy range, but the fine structure that the
experiment shows is not reproduced. A CI calculation \cite{koutecky96} based
on isomer (b) reproduced some of the experimentally observed patterns, but
also could not explain all the peaks, and recent three-dimensional TDLDA
calculations \cite{pgrsuraudtobepu} have lead to similar results as
the collective 
model. An explanation of all details in the experimental spectrum therefore
might require to consider a mixture of low-energy structures, and this will be
discussed in detail in a separate publication.\cite{pgrsuraudtobepu}

\begin{figure}[htb]
  \PostScript{0}{6}{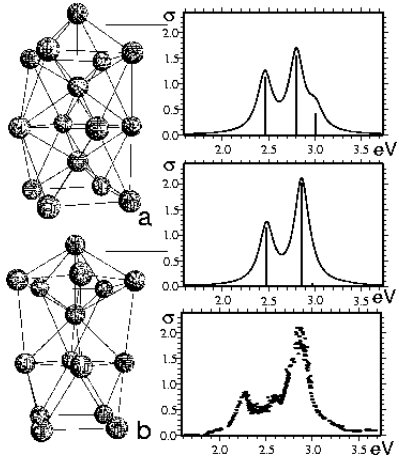}
\caption{Low-energy configurations and photoabsorption spectra for 
  $\mathrm Na_{15}^+$. A phenomenological Lorentzian line broadening
  of width 0.2 eV has been applied. Dotted curve: experimental cross
  section.\cite{haberlandepjd}}
\label{na15pall}
\end{figure}
For $\mathrm{Na}_{15}^+$ CAPS finds four low-energy structures. The
two lowest ones, (a) and (b) in Fig.\ \ref{na15pall}, are very close
in energy in both CAPS and the three-dimensional calculations. (a)
corresponds to a distorted $\mathrm{Na}_{11}^+$ (a) with four atoms
added along the $z$-axis, and (b) can be understood as a pentagonal
bipyramid sharing one ion with $\mathrm{Na}_{9}^+$ (a). Structure (c)
results from structure (a) by moving one of the inner single atoms to
the bottom face, and the oblate (d) corresponds to a
configuration found\cite{spiegelmann2,epjd} for $\mathrm{Na}_{14}$,
but with one atom added 
along the symmetry axis. In agreement with the experimental
photoabsorption spectrum, our results point to 
prolate clusters as the relevant structures and thus again verify the
prediction of the deformed, structure averaged jellium model
(SAJM).\cite{bmprb,sajm}  

\begin{figure}[hbt]
  \PostScript{0}{5}{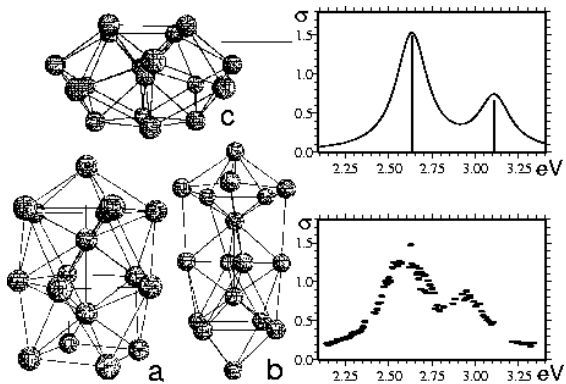}
\caption{Low-energy configurations for 
  $\mathrm Na_{17}^+$, and photoabsorption spectrum for isomer (c). A
  phenomenological Lorentzian line broadening of width 0.22 eV has
  been applied. Dotted curve: experimental cross
  section.\cite{haberlandepjd}}
\label{na17pall}
\end{figure}

The ground state structure found for $\mathrm{Na}_{17}^+$, (a) in
Fig.\ \ref{na17pall}, is close to the one found in the extended
H\"uckel model for the neutral cluster.\cite{spiegelmann2} It
consists of an icosahedral core with a crown of four atoms, and the
only difference to the result of Ref.\ \onlinecite{spiegelmann2} is that the
cylindrical averaging forces the crown atoms into a square. This shows
that for such sizes the differences between charged and neutral
clusters can already be small. Structure (b), which is slightly higher
in energy than (a) in the three-dimensional calculations, has an even
more prolate valence electron density, and the collective resonances
for these clusters lie at 2.59 eV and 2.86 eV [structure (a)] and at
2.34 eV and 2.95 eV [structure (b)].  They are thus in contrast to the
experimental spectrum that points to an oblate cluster. CAPS finds one
more isomer, structure (c), that indeed leads to an oblate valence
electron density. 
In CAPS, it is 0.16 eV higher than (a), with a half
occupied orbital. 
The collective spectrum for this isomer is very
close to the experimental one, as shown in Fig.\
\ref{na17pall}. However, also the three-dimensional calculation gives
a non-negligible energy difference between (c) and the prolate
isomers, cf.\ Table \ref{allen}. Therefore, in this case a
three-dimensional relaxation of the ions would be necessary to check
whether Jahn-Teller distortions lower the energy of isomer (c) so much
that it can account for the experimental spectrum.

\begin{figure}[htb]
  \PostScript{0}{9}{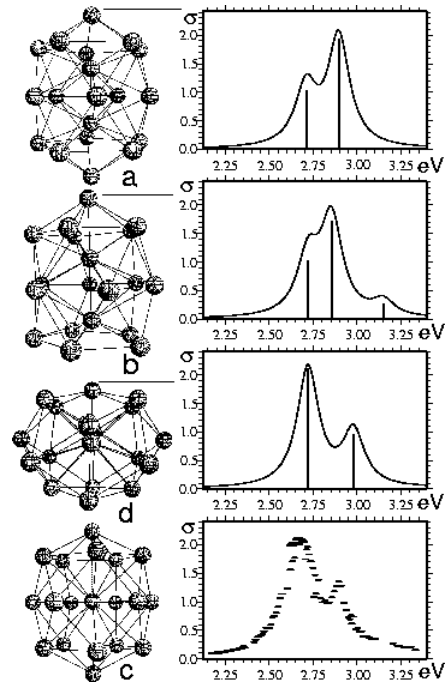}
\caption{Low-energy configurations and photoabsorption spectra for 
  $\mathrm Na_{19}^+$.  A phenomenological Lorentzian line broadening
  of width 0.15 eV has been applied. Dotted curve: experimental cross
  section.\cite{haberlandepjd}}
\label{na19pall}
\end{figure}

For $\mathrm{Na}_{19}^+$ CAPS finds the four stable geometries shown
in Fig.\ \ref{na19pall}. The double-icosahedron (a) is the ground
state, and it is separated from structure (b) by 0.23 eV, from (c) by
0.30 eV, and from (d) by 0.35 eV. However, in the calculations without
axial restriction, the energetic differences are considerably reduced:
isomer (d) becomes equivalent to isomer (b), and both are separated
from the ground state by only 0.12 eV. The experimental
photoabsorption spectrum sheds further light on the situation. It
shows a high peak at about 2.7 eV, followed by a lower one at about
2.9 eV, and is thus very similar to the collective spectrum of isomer
(d). The spectra of the other isomers all give the peaks in reversed
order and thus do not resemble the experiment. (The collective
spectrum of (c), which is not shown in order not to overload Fig.\
\ref{na19pall}, shows a small peak at 2.84 eV and a higher one at
2.93 eV.)  Since an internal cluster temperature of about 60 K is
sufficient to go 
from isomer (a) to (d), and since the temperature in the experiment is
about 105 K, 
it is plausible that isomer (d) will contribute to the
spectrum.  The experimental spectrum however looks as if it is
dominated by isomer (d).  One explanation for this can be that
structure (d) has a larger ``catchment area'' than (a) and is
entropically favored. The second possibility, which is 
suggested by the relatively large difference between the energy found
in CAPS and the three-dimensional calculation, is that a
three-dimensional relaxation of the ions would turn structure (d) into
the ground state.

\begin{figure}[htb]
  \PostScript{0}{6}{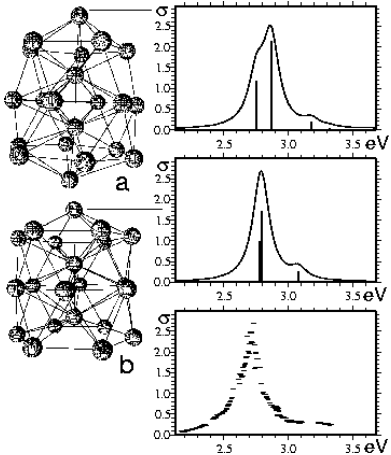}
\caption{Low-energy configurations and photoabsorption spectra for 
  $\mathrm Na_{21}^+$. A phenomenological Lorentzian line broadening
  of width 0.15 eV has been applied. Dotted curve: experimental cross
  section.\cite{expcoldclosed}}
\label{na21pall}
\end{figure}
Two degenerate structures are found for $\mathrm{Na}_{21}^+$ in CAPS,
and also the three-dimensional calculation leads to nearly equal
energies. Structure (a) results from the double icosahedron of
$\mathrm{Na}_{19}^+$ (a) by adding one ion to each of the lower tow
pentagons, and (b) results from (a) by moving the lowest ion to the
uppermost pentagon. Both geometries lead to collective spectra with
two main transitions and some strength above the main resonances.
This is in agreement with the experimental data
\cite{expcoldclosed,haberlandepjd} 
that also show this high-energy tail.  Concerning the main
transitions, the two isomers are somewhat different: in structure (a)
the main peaks are separated by about 0.1 eV, whereas they nearly
fall together for structure (b). Since the structures are extremely
close in total energy and since the main peaks of (b) energetically
fall exactly between the main peaks of (a), the experimental spectrum
can consistently be explained as resulting from a mixture of both
isomers.

\begin{figure}[htb]
  \PostScript{0}{4}{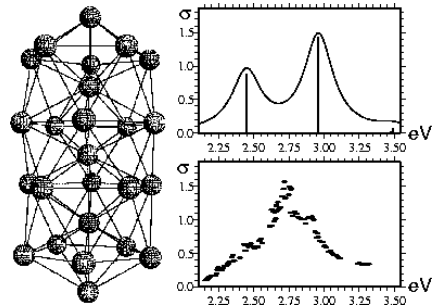}
\caption{CAPS ground-state configuration and photoabsorption spectrum for
  $\mathrm Na_{25}^+$. A phenomenological Lorentzian line broadening
  of width 0.25 eV has been applied. Dotted curve: experimental cross
  section.\cite{haberlandepjd}}
\label{na25pall}
\end{figure}
The simulations performed for $\mathrm{Na}_{25}^+$ led to strongly
prolate shapes. This is partially understandable on the basis of the
SAJM: \cite{sajm} the valence electrons force the cluster into the
prolate regime. But the inclusion of ionic structure even enhances the
prolate tendency, because the lowest configuration, shown in Fig.\
\ref{na25pall}, is a triple icosahedron, and in order to build this
configuration, a quadrupole moment larger than the jellium prediction
is necessary. From this observation it also becomes clear that the
pentagonal bipyramid which has been observed as a building block of
the smaller clusters\cite{kouteckyrev,roethlis,spiegelmann2} is also
important for the larger sizes. For the 
triple icosahedron, collective resonances are observed at 2.44 eV and
2.96 eV, and also the experiments \cite{borggreen1,haberlandepjd}
indicate peaks at these energies. But the latter shows an additional
peak around 2.75 eV which is not found in the calculations.
Comparing this to the situation encountered\cite{prbrc} for
$\mathrm{Na}_{27}^+$ shows that the geometries of the two 
clusters are closely related: $\mathrm{Na}_{27}^+$ is the triple
icosahedron with two atoms added to the central pentagons. This shows
the consistency of the structure calculations, and for
$\mathrm{Na}_{27}^+$ the collective spectrum nicely matches with the
experimental data. The triple icosahedron is energetically strongly
favored over the other structures that CAPS finds. All of them are
variations of the ground state geometry with local distortions, and a
three-dimensional calculation was performed for the structure that in
CAPS is the second lowest. Here, the difference in total energy is a
little smaller than in CAPS, but still large, cf.\ Table
\ref{allen}. Since furthermore all low-energy structures give rise to
collective spectra that are similar to the one shown in Fig.\
\ref{na25pall}, the explanation for the middle peak observed for
$\mathrm{Na}_{25}^+$ remains an open question.

\begin{figure}[htb]
  \PostScript{0}{3}{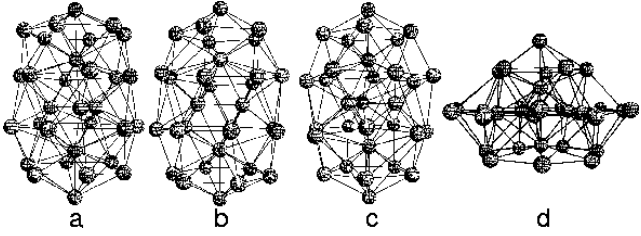}
\caption{Low-energy configurations for $\mathrm Na_{31}^+$.}
\label{struct31pall}
\end{figure}

\begin{figure}[htb]
  \PostScript{0}{5}{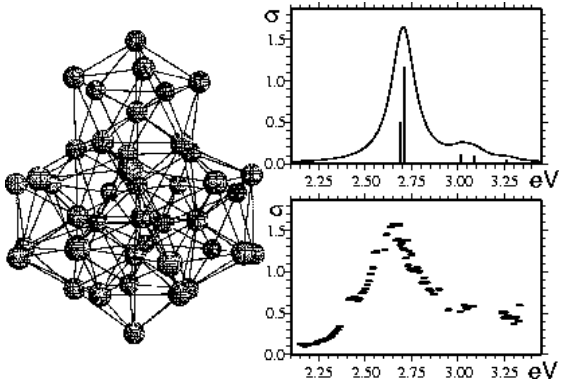}
\caption{CAPS ground state configuration and photoabsorption spectrum for 
  $\mathrm Na_{43}^+$. A phenomenological Lorentzian line broadening
  of width 0.15 eV has been applied. Dotted curve: experimental cross
  section.\cite{haberlandepjd}}
\label{na43pall}
\end{figure}

Comparing the structures of $\mathrm Na_{25}^+$ and $\mathrm
Na_{27}^+$ to the ones found for $\mathrm Na_{31}^+$, $\mathrm
Na_{41}^+$ and $\mathrm 
Na_{43}^+$ (Fig.\ \ref{struct31pall}, Fig.\ \ref{na43pall}, and Refs.\
\onlinecite{rytkoenen,prbrc}) allows 
to identify a growth pattern: the triple icosahedron serves as a
building block for the larger clusters. Additional atoms are added on
outside faces, as seen in the ground states of $\mathrm Na_{27}^+$ and
$\mathrm Na_{43}^+$, or can be packed between the pentagonal subunits,
as observed for the isomers of $\mathrm Na_{27}^+$, where the two
additional ions are placed at different positions ``inside'' the
$\mathrm Na_{25}^+$ structure. 
The collective spectra of the low-energy structures (a) -- (c) of
$\mathrm{Na}_{31}^+$ reflect the prolate electron densities of these
clusters and show one peak around 2.6 eV, carrying about 30\%
oscillator strength, another one around 2.95 eV carrying about 50\%
strength, and the rest of the strength scattered at higher
energies. Structure (d) is not stable against Jahn Teller distortions
in three dimensions. In view of the cluster size, the CAPS differences
in total energy for $\mathrm{Na}_{31}^+$ are rather small, and they
are even smaller in the three-dimensional approach. Since the
temperature in the measurement of the photoabsorption cross section
was again at least 105 K, it can be concluded from the calculations
that a variety of isomers will contribute to the experimental
spectrum. It is thus not astonishing that the available experimental
data \cite{borggreen1,haberlandepjd} do not resolve separated peaks but
show a rather broad bump.

The pronounced deformation seen for $\mathrm Na_{43}^+$ in Fig.\
\ref{na43pall} results from an interplay between ions and valence
electrons. The electrons ``push'' the cluster into the octupole, i.e.\
pear-shaped form, and the ions arrange under this ``constraint''. But
the ions favor the icosahedral core, and this increases the
deformation. The octupole moment therefore is larger by a factor 1.15
than in the SAJM.\cite{bmprb} The
photoabsorption spectrum calculated for $\mathrm Na_{43}^+$ shows a
strong peak at 2.7 eV, followed by a high-energy tail. It is in close
agreement with the experimental data, which thus support our structure
calculations.

\begin{figure}[htb]
  \PostScript{0}{11}{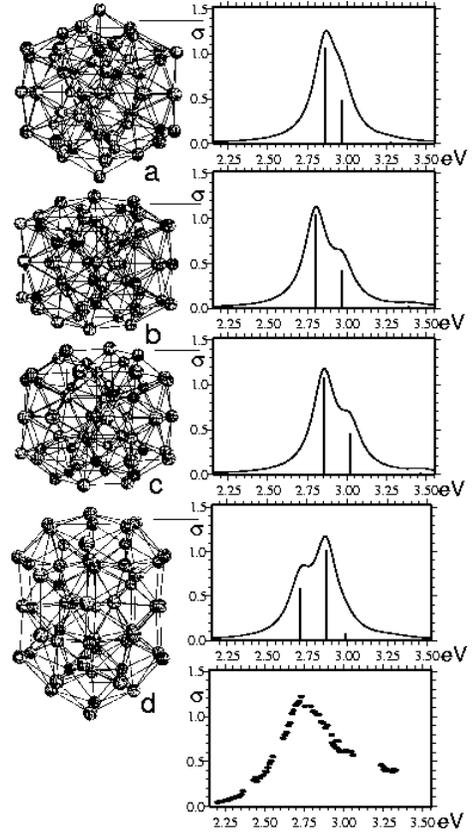}
\caption{Low-energy configurations and corresponding photoabsorption
  spectra for $\mathrm Na_{55}^+$. A phenomenological Lorentzian line
  broadening of width 0.17 eV has been applied. Dotted curve:
  experimental cross section.\cite{haberlandepjd} See text for a discussion
  of further experiments.\cite{meibom}}
\label{na55psome}
\end{figure}
Based on the shell and deformed jellium model a configuration would be
expected for 
$\mathrm{Na}_{55}^+$ that gives rise to a prolate valence electron
density. The CAPS calculations, however, consistently led to nearly
spherical or slightly oblate clusters as low-energy
configurations. The lowest energy was found for structure (a) in Fig.\
\ref{na55psome}. It has a fivefold symmetry axis and again shows
the close-to-icosahedral core discussed previously. Within 0.4 eV
structures (b) and (c) are found with valence electron densities that
are more oblate. These structures show half-occupied highest orbitals,
indicating that they would Jahn-Teller relax if the axial restriction
on the electrons were lifted. The lowest prolate isomer that was
found, structure (d), has a quadrupole moment that is close to the one
predicted by the SAJM \cite{sajm} but in CAPS this configuration is
0.73 eV higher than the ground state. The annealing was also started
from an icosahedron with the nearest-neighbor distance of bulk
sodium. In this calculation, the strictly geometrical icosahedron
transforms into structure (a), which can be regarded as a slightly
distorted icosahedron. The experimental photoabsorption spectrum of
$\mathrm{Na}_{55}^+$ as measured by the Freiburg group is shown in the
lowest panel of Fig.\ \ref{na55psome}, and the spectrum has also
been measured by Meibom {\it et al.}\cite{meibom} Both spectra have
in common that one broad peak is seen, followed by a second, smaller
peak or a high-energy shoulder. This is in contrast to the prediction
of the jellium model, because the prolate structures found there give
the peaks in reversed order, similar to the spectrum of isomer
(d). The CAPS low-energy structures, however, lead to collective
spectra with a high peak followed by a lower one, and are thus in
better agreement with the experiment. The total oscillator strength
measured in the experiment \cite{meibom} was 70\% - 80\%.  This also
agrees with the collective model calculations that find 79\% strength
within the plotted range for structure (a), 76\% for (b) and 80\% for
(c). As a test for the collective model, a TDLDA calculation with
excitation in z-direction was performed for structure (a). In the
TDLDA, the high-energy shoulder is more 
pronounced than in the collective model. To further clear up the
situation, the structures (a), (b) and (d) were relaxed in
Born-Oppenheimer {\it ab initio} molecular dynamics,\cite{ldamd}
i.e., fully three-dimensional with the Troullier-Martins
pseudopotential.\cite{jaakkopriv} The bond lengths of all three structures
shrinked by a few percent due to the different pseudopotential, and
isomers (b) and (d) distorted slightly, but
otherwise the geometries stayed unaltered. As seen in Table
\ref{allen}, the differences in total energy are somewhat smaller in
the {\it ab initio} calculations, but the ordering of isomers is the
same as in CAPS. Finally, as reported in Ref.\ \onlinecite{akola00}, the CAPS
structure (a) was annealed for 10 ps at about 220 K. In this
annealing, the cluster became even more similar to an icosahedron and
its overall shape oscillated around the nearly spherical shape of the
icosahedron. Thus, the {\it ab initio} calculations confirm the
finding that in contrast to the deformed jellium model prediction,
$\mathrm{Na}_{55}^+$ at low to moderate temperatures has a close to spherical
valence electron density.

\begin{figure}[htb]
  \PostScript{0}{9}{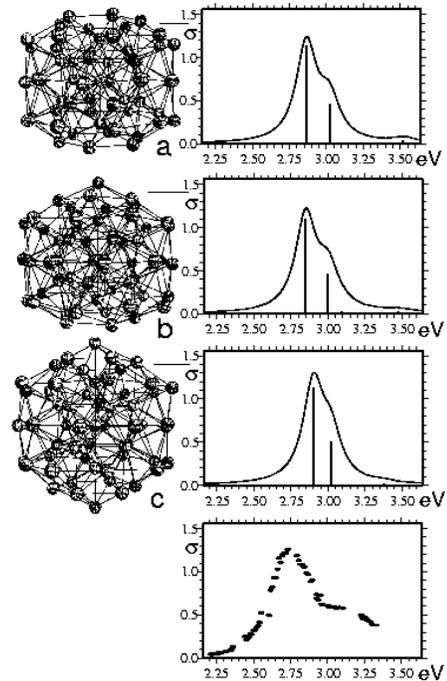}
\caption{Low-energy configurations and corresponding photoabsorption
  spectra for $\mathrm Na_{57}^+$. A phenomenological Lorentzian line
  broadening of width 0.17 eV has been applied. Dotted curve:
  experimental cross section.\cite{haberlandepjd} See text for a discussion
  of further experiments.\cite{meibom}}
\label{na57psome}
\end{figure}
The CAPS results for $\mathrm{Na}_{57}^+$, see
Fig. \ref{na57psome}, are consistent with the results for
$\mathrm{Na}_{55}^+$: again the low-energy structures are not prolate,
but spherical or slightly oblate. The collective spectra somewhat
underestimate the high-energy shoulder, and this can be attributed to
selective particle-hole effects just as in the case of
$\mathrm{Na}_{55}^+$. But the overall agreement with experiment
\cite{meibom,haberlandepjd} is considerably better than in the
deformed-jellium calculation. Structure (a) has the lowest energy,
followed by isomers (b) and (c) that are higher by 0.22 eV.  For
structure (c) we observe a half-occupied orbital. With respect to the
ionic geometry, structure (a) is similar to the third isomer of
$\mathrm{Na}_{55}^+$, and (c) corresponds to the ground state of
$\mathrm{Na}_{55}^+$ with the two additional ions added on top and
bottom faces. Another isomer, not shown in Fig.\ \ref{na57psome}
for the sake of clarity, is found, and there the two additional ions
are added in the equatorial plane of $\mathrm{Na}_{55}^+$ (a).

In the case of $\mathrm{Na}_{59}^+$ the search for low-energy
structures has not been as exhaustive as for the other clusters. But
several annealing runs were started from random configurations, and
the low-energy geometries that were found again showed the ions
arranged rather regularly and similar to the structures just
discussed. Therefore, further simulations were started from the
geometries found for $\mathrm{Na}_{55}^+$ and $\mathrm{Na}_{57}^+$,
plus four or two atoms, respectively, that were added at randomly
chosen sites. One of these runs led to a geometry that corresponds to
$\mathrm{Na}_{57}^+$ (a), capped by two atoms on top and bottom. This
structure has the lowest energy of all that were found, and it does
not show signs of Jahn-Teller instability. The comparison between the
experimental \cite{haberlandepjd} and theoretical photoabsorption
spectra fits into the previous discussion. The strongest collective
resonance is seen at 2.8 eV, followed by a smaller one at 2.9 eV and a
little strength scattered around 3.25 eV, i.e., the collective model
leads to qualitative agreement with the experimental data.

\begin{table}[t]
\begin{tabular}{|c|c|c|}
\hline 
 Cluster & $\Delta E_{\mathrm CAPS}$/eV  & $\Delta E_{\mathrm 3D}$/eV \\
\hline
\hline
$\mathrm{Na}_9^+$ a &    G &    G \\
\hline
$\mathrm{Na}_9^+$ b & 0.10 & 0.12 \\
\hline
$\mathrm{Na}_9^+$ c & 0.23 &   G' \\
\hline
\hline
$\mathrm{Na}_{11}^+$ a &   G  & 0.16 \\
\hline
$\mathrm{Na}_{11}^+$ b & 0.11 &    G \\
\hline
$\mathrm{Na}_{11}^+$ c & 0.19 & 0.10 \\
\hline
\hline
$\mathrm{Na}_{15}^+$ a &   G  & 0.04 \\
\hline
$\mathrm{Na}_{15}^+$ b & 0.03 &    G \\
\hline
$\mathrm{Na}_{15}^+$ c & 0.15 & 0.16 \\
\hline
$\mathrm{Na}_{15}^+$ d & 0.23 & 0.15 \\
\hline
\hline
$\mathrm{Na}_{17}^+$ a &   G  &    G \\
\hline
$\mathrm{Na}_{17}^+$ b &   G' & 0.08 \\
\hline
$\mathrm{Na}_{17}^+$ c & 0.16 & 0.20 \\
\hline
\hline
$\mathrm{Na}_{19}^+$ a &   G  &    G \\
\hline
$\mathrm{Na}_{19}^+$ b & 0.23 & 0.12 \\
\hline
$\mathrm{Na}_{19}^+$ c & 0.30 & 0.20 \\
\hline
$\mathrm{Na}_{19}^+$ d & 0.35 & 0.12 \\
\hline
\hline
$\mathrm{Na}_{21}^+$ a &   G  &    G \\
\hline
$\mathrm{Na}_{21}^+$ b &   G' & 0.03 \\
\hline
\hline
$\mathrm{Na}_{25}^+$ a &   G  &    G \\
\hline
$\mathrm{Na}_{25}^+$ b & 0.41 & 0.29 \\
\hline
\hline
$\mathrm{Na}_{31}^+$ a &   G  &    G \\
\hline
$\mathrm{Na}_{31}^+$ b & 0.14 & 0.08 \\
\hline
$\mathrm{Na}_{31}^+$ c & 0.19 & 0.10 \\
\hline
\hline
$\mathrm{Na}_{55}^+$ a &   G  &    G \\
\hline
$\mathrm{Na}_{55}^+$ b & 0.41 & 0.30 \\
\hline
$\mathrm{Na}_{55}^+$ d & 0.73 & 0.31 \\
\hline
\end{tabular}
\caption{Differences in total energy for $\mathrm{Na}_N^+$ clusters. Small
  letters behind the cluster symbol label structures and refer to
  Fig.\ \ref{na9pall} -- Fig.\ \ref{na55psome} and the main text. G
  (and G') denote 
  the structure with lowest energy for a given cluster size. The left
  column for each size gives the difference in total energy between
  the corresponding structure and G as found in CAPS. The right column
  gives the energetic differences found for the same structures in the
  three-dimensional Kohn-Sham calculation. The 3D values for
  $\mathrm{Na}_{55}^+$ were obtained by relaxing the CAPS structures
  in {\it ab initio} Born-Oppenheimer molecular
  dynamics,\cite{jaakkopriv} see text for discussion.} 
\label{allen}
\end{table}

\section{Discussion and conclusion}
\label{conclude}

The systematic survey showed that for most of the smaller Na clusters,
the overall deformation is determined by electronic shell effects even
when the ionic structure is explicitly included. This explains the
great success of the deformed jellium model. However, the survey at
the same time clearly exhibits the limitations of the jellium
picture. Besides the fact that some
spectroscopic patterns, e.g., 
as seen for $\mathrm{Na}_9^+$, are directly related to
details in the ionic configuration, the fundamental
improvement brought about by taking into account the ionic structure
is that growth systematics can be identified. In the smallest
clusters, the pentagonal bipyramid is a frequently appearing structure
and is seen, e.g., in $\mathrm{Na}_{7}^+$, $\mathrm{Na}_{15}^+$,
$\mathrm{Na}_{17}^+$. In $\mathrm{Na}_{19}^+$, three of these
smallest units are combined to build up the double icosahedron, and
from then on, the electronic shell effects and the preference of the
ionic structure for icosahedral geometries work hand in hand in
determining the cluster structure. This is seen most impressively for
$\mathrm{Na}_{43}^+$. At $\mathrm{Na}_{55}^+$, however, the situation
changes. Due to the ionic shell closing, the influence of the ions wins over
the electronic shell effects, resulting in nearly spherical
structures. This finding is supported by the experimental
photoabsorption data and has been verified in {\it ab initio}
calculations.

In summary, we have presented a local pseudopotential for sodium that
accurately 
models the core-valence interaction and correctly describes the atom,
the bulk, and finite clusters. We have demonstrated the influence of
pseudopotentials on structural properties and the direct influence of
the bond lengths on the resonance positions. This shows that even for the
most simple metal, a pseudopotential must be constructed carefully.
Cluster structures were calculated in axially averaged and
three-dimensional Kohn-Sham calculations. A detailed comparison with
{\it ab initio} work demonstrated the validity of the CAPS as a tool
for the approximate determination of cluster configurations. The
systematical survey for clusters with up to 59 ions revealed an
icosahedral growth pattern. Collective resonances were calculated for
the theoretically determined structures, and comparison with the
experimental photoabsorption spectra confirmed the results of our
structure calculations.  Thus, a distinct step in the growing process
of matter has been put into evidence, namely the transition from
electronically to ionically determined configurations.

\acknowledgments

We thank J.\ Akola for the communication of his results prior to publication,
and one of us (S. K\"ummel) further acknowledges stimulating
discussions with A.\ Aguado and F.\ Calvo. This work has been
partially supported by the Deutsche Forschungsgemeinschaft.


\end{document}